\documentclass[12pt]{article}
\usepackage{authblk}
\usepackage{graphicx}
\usepackage{url}
\usepackage{hyperref}
\usepackage[margin=1in]{geometry}

\usepackage{amsmath}
\usepackage{braket}

\DeclareMathOperator{\Tr}{Tr}

\newcommand{\D}[2]{\frac{\partial #1}{\partial #2}}

\renewcommand{\vec}[1]{\boldsymbol{#1}}
\newcommand{\F}{\mathcal{F}}

\newcommand{\I}{\mathcal{I}}
\newcommand{\rp}{r_+}
\newcommand{\rn}{r_-}

\title{Thermal Fisher information for a rotating BTZ black hole}
\author{Everett A. Patterson}
\author{Robert B. Mann}
\affil{Department of Physics and Astronomy, University of Waterloo, Waterloo, Ontario, Canada N2L 3G1}

\begin{document}

% \title{Thermal Fisher information for a rotating BTZ black hole}

% \author{Everett A. Patterson and Robert B. Mann}

\maketitle

\abstract{Relativistic quantum metrology provides a framework within which we can quantify the quality of measurement and estimation procedures while accounting for both quantum and relativistic effects. The chief measure for describing such procedures is the Fisher information, which quantifies how sensitive a given estimation is to a variance of some underlying parameter. Recently, the Fisher information has been used to quantify the spacetime information accessible to two-level quantum particle detectors.
We have previously shown that such a system is capable of discerning black hole mass for static black holes in 2+1 dimensions. 
Here, we extend these results to the astrophysically interesting case of rotating black holes and show that the Fisher information is also sensitive to the rotation of a black hole.}

%%%%%%%%%%%%%%%%%%%%%%%%%%%%%%%%%%%%%%%%%%

\section{Introduction}

Since their inception at the beginning of the 20th century, general relativity (GR) and quantum theory have cemented their identity as the cornerstones of modern physics. Supported by the strongest of experimental evidence \cite{Nature_2019_GR,FMSG_2023_SMtest}, Einstein's GR and quantum field theory (QFT) are often touted as the two most successful scientific theories. However, for all their predictive power within their respective regimes of validity, an overarching theory of quantum gravity remains elusives

% Indeed, the development of a complete theory of quantum gravity remains one of the most important open problems in physics. Despite serious efforts developing such a theory from the top-down, both the string and loop approaches seem to have stalled in their quest to describe our universe.
Here, we consider %instead 
a semi-classical approach to quantum gravity using quantum field theory in curved spacetime \cite{BirrellDavies_1984}.
While this approach does not purport itself to be a complete theory of quantum gravity, it allows us to faithfully examine interactions between quantum and relativistic effects in the regimes of low-energy quantum fields (i.e., neglecting backreaction). 
In particular, this approach has provided us with the framework to describe the Unruh effect \cite{Unruh_1976_Notes-on-BH-evaporation} %, 
and the Hawking radiation emitted by black holes \cite{Hawking_1975_Particle-creation-by-BH}. 
The notion of spacetime temperature that arises from these examples is an important feature at the interface between quantum and gravitational effects, and is what we will set out to better understand in this paper.

In order to make these sorts of properties operationally accessible, it is common to employ particle detectors, such as the Unruh-DeWitt particle detector \cite{Unruh_1976_Notes-on-BH-evaporation,DeWitt_1979}. By locally coupling such a detector to the quantum field, it is possible to gain information about global properties of the quantum field and the underlying spacetime by performing measurements on the particle detector. 
These types of protocols are frequently described in the context of relativistic quantum information (RQI) \cite{MannRalph_2012_RQI}.

In general, RQI provides us with a framework in which to leverage tools and knowledge from quantum information theory to relativistic applications.
In practice, much of RQI concerns itself with considering how relativistic effects might alter quantum information protocols \cite{Dowker_2011_Useless,Sorkin_1993_Impossible} or allow for new ones (e.g., quantum energy teleportation \cite{Hotta_2008}, entanglement harvesting \cite{PKMM_2015_HarvestingCorrelations, PKMM_2016_EntanglementHarvesting}, relativistic quantum communication channels \cite{ClicheKempf_2010_RQCC, Landulfo_2016_NonperturbativeRQCC,Kasprzak-Tjoa_2024_Transmission-QI-QFTCS}). 
A natural extension of the latter asks how quantum information protocols might be used to probe underlying relativistic structures \cite{GKMTT_2021_QuantumImprints,PercheMM_2022_Geometry}.

Closely related is the field of relativistic quantum metrology, which examines questions surrounding measurement in the presence of relativistic and quantum effects \cite{Ahmadi-BSAF_2014_RQM,Ahmadi-BF_2014_qm4rqf}.
It is in effect an extension of standard quantum metrology \cite{Braunstein-Caves_1994, Giovannetti_2006_Q-Metrology, Paris_2009_qe4qt, Giovannetti_2011_Adv-Q-Metrology}, which has found important applications to quantum sensing \cite{Degen_2016_Quantum-sensing, DeMille_2024}, most notably with modern interferometric gravitational wave detection experiment \cite{LIGO_2011, KAGRA_2020}.
Further accounting for relativity has opened the door to further improving metrological procedures \cite{Cepollaro_2022, Zhao-Yang-Chiribella_2020_qm4ico}.
In much of quantum metrology, the Fisher information serves as the measure by which we can quantify the relevant metrological value.

In the context of RQI, quantum metrology has been used to estimate the Hubble parameter \cite{HFZF_2018_Q-estimation-exp-spacetime} and conical deficits in cosmic string spacetimes \cite{Yang-WWJI_2020_pe-cosmic-string, YJT_2022_Cosmic-string-pe-EM}.
It has also been used to estimate the thermal properties of the quantum field by using quantum detectors in spacetime \cite{Aspachs-AF_2010_qe4Unruh, FengZhang_2021_QFI-Unruh, Tian-WJD_2016_Thermometry-Unruh, YZJJ_2022_Q-thermometry-cosmic-string}.
More specifically, quantum metrology has been applied to estimating the temperature of the quantum field, as quantified by the thermal Fisher information, in asymptotically curved spacetimes \cite{DuMann_2021}, where it can discern differences between the global topologies of de Sitter (dS) and anti-de Sitter (AdS) spacetimes.
The thermal Fisher information has also been shown to distinguish between boundary conditions and masses of a static black hole spacetime in 2+1 dimensions \cite{PattersonMann_2023}.

Here, we extend these recent results to the case of a rotating BTZ spacetime, where the existence of angular momentum is known to lead to interesting variations for RQI results.
For example, the entanglement harvested by a pair of quantum detectors is amplified by the presence of angular momentum in a BTZ black hole, with the effect being most pronounced for near-extremal black holes \cite{Robbins_2020_Ent-amp-rBTZ}. 
There is also work describing the effect of rotation on the anti-Hawking phenomena for BTZ spacetimes \cite{Robbins_2021_Anti-Hawking-rBTZ}. There, it was shown that the weak anti-Hawking effect (dependent on the detector's response) is accentuated by rotation in general, while the strong anti-Hawking effect (dependent on the detector's perceived temperature) can display antithetical behaviour depending on the boundary condition.

\section{General Formalism}\label{sec:formalism}

In this section, we present the tools and formalism employed in the rest of this paper. We begin by describing the spacetime under consideration in section \ref{sec:spacetime}. Then, we will present the Unruh-DeWitt particle detector in section \ref{sec:UDW-detector}, before defining the Fisher information in section \ref{sec:FI}. 
We will employ natural units of $c=\hbar=G=k_B=1$.

\subsection{Spacetime: The Rotating BTZ Black Hole}\label{sec:spacetime}

When restricted to 2+1 dimensions, the Einstein equations admit a single black hole solution, known as the BTZ black hole \cite{BTZ_1992}. In contrast to the standard 3+1 dimensional Schwarzchild black hole, the black hole in 2+1 dimensions requires a negative cosmological constant, $\Lambda$, in order to be well-defined. As a result, the BTZ black hole is closely related to anti-de Sitter (AdS) spacetime and is thus commonly investigated in the context of AdS-CFT holography.

In following with the no-hair conjecture for classical black holes, the uncharged BTZ black hole is fully described by only two parameters: a unitless mass $M$ and an angular momentum $J$. These parameters, can be described in terms of the inner and outer radii parameters $\rn$ and $\rp$ as
\begin{equation}
\label{eq:M-J-vs-rp-rn}
    M = \frac{\rp^2+\rn^2}{\ell^2}, \qquad J = \frac{2\rp\rn}{\ell},
\end{equation}
where $\ell$ is the AdS length scale related to cosmological constant by $\Lambda = -1/\ell^2$. It will sometimes be mathematically convenient to express equations in terms of the radii, however we will describe all our physical interpretations in terms of the mass $M$ and angular momentum $J$.

The line element describing the rotating BTZ black hole spacetime is given by 
\begin{equation}
\label{eq:rotating-btz-metric}
    ds^2 = - \left( N^\perp \right)^2 dt^2 + f^{-2} dr^2 + r^2 \left( d\phi + N^\phi dt \right)^2,
\end{equation}
where, $N^\perp = f = \sqrt{-M +\frac{r^2}{\ell^2} +\frac{J^2}{4r^2}}$ is the lapse function and $N^\phi = -\frac{J}{2r^2}$ is the angular shift function. 

It is worth noting that the angular momentum of a black hole is bounded from above by its mass. Expressed as $|J| \leq M\ell$, we say that a rotating black hole is extremal when the equality is satisfied. Expressed in terms of the inner and outer radii, we have that a black hole is extremal when the two radii are degenerate with $\rp=\rn$.

On the other hand, when the black hole's has zero angular momentum, we recover the static BTZ black hole spacetime. Mathematically, this corresponds to $J=0$ or equivalently $\rn=0$, while the remaining radius persists as the black hole horizon radius.

\subsubsection{Co-Rotating detector trajectory}

In this paper, we want to examine the thermal response of an observer (more precisely a quantum detector, which we will define in Section \ref{sec:UDW-detector}) in the presence of the rotating black hole.

We will restrict ourselves to the case of a co-rotating detector trajectory for a few reasons: First, this trajectory is physically meaningful because it allows us to focus on the rotation of the black hole itself, rather than its rotation relative to the detector. Moreover, this trajectory has been the one considered in previous RQI literature concerning rotating BTZ black holes \cite{Robbins_2020_Ent-amp-rBTZ,Robbins_2021_Anti-Hawking-rBTZ} and results in static detector outside a static black hole in the limit $J\to0$.
Lastly, it is computationally challenging to compute the response of a static detector outside a rotating black hole.
As such, the detector under consideration will remain at a fixed radius away from the center of the black hole, but will be `orbiting' around the black hole with an angular velocity equal to that of the black hole.

In order to succinctly describe the trajectory of our co-rotating detector, we will employ the inner and outer radii, recalling their relation to the mass and angular momentum given by Equation (\ref{eq:M-J-vs-rp-rn}). The trajectory in the co-rotating frame is given by 
\begin{equation}
    \left\{ r=R, \quad t= \frac{\ell\tau}{\gamma}, \quad \phi=\frac{\rn\tau}{\rp\gamma} \right\},
\end{equation}
where $\tau$ is the time in the co-rotating frame, and $\gamma = \sqrt{(r^2-r_+^2)(r_+^2-r_-^2)}/r_+$ is the redshift factor at the radial position $r$.
Note that we can also relate the angular and time coordinates by $\phi = \frac{r_-}{\ell r_+} t$

\subsubsection{QFT in curved spacetime}

In order to talk about the thermal and quantum behaviours associated with black holes, we must be working in a framework that allows for both quantum and gravitational effects. While a full theory of quantum gravity remains elusive, we are able obtain approximate results in the framework of quantum field theory in curved spacetime (QFTCS) \cite{BirrellDavies_1984}. QFTCS serves as an approximation to a theory of quantum gravity in the regime of weak (low energy) quantum fields. Here we assume that the curvature of spacetime will have an effect on the quantum fields, but that the fields themselves do not back-react onto the spacetime.

An important part of many applications of QFTCS in the context of RQI is that the properties of the QFT are fully encoded by the $n$-point correlation functions \cite{PercheMM_2022_Geometry}. 
Moreover, when considering the special case of physically-relevant quasifree states, these $n$-point correlation functions are entirely determined by the vacuum 2-point correlation function, also known as the Wightman function.

Examples of quasifree state include
(squeezed)
vacuum and thermal states. In particular, the Hartle-Hawking vacuum state which we will be considering here is a quasifree state.

In the case of the BTZ spacetime, this Hartle-Hawking vacuum Wightman function can be expressed as an image sum \cite{Carlip_1995_BTZ,Lifschytz-Ortiz_1993}:
\begin{align}
\label{eq:rotating-btz-wightman}
    W_{\rm {BTZ}}(x, x') = \frac{1}{4 \pi \sqrt{2} \ell } \sum_{n=-\infty}^\infty  \left[ \frac{1}{\sqrt{\sigma_n}} -  \frac{\zeta}{\sqrt{\sigma_n + 2}} \right],
\end{align}
with $\zeta=\{-1,0,1\}$ corresponding to Neumann, transparent, and Dirichlet boundary conditions, and the square of the distance between the points under examination being given by
\begin{align}
\label{eq:rbtz-squared-distance}
    \sigma_n(x,x') =& -1 + \sqrt{\alpha(r)\alpha(r')} \cosh\! \left[\frac{r_+}{\ell} ( \Delta \phi - 2 \pi n) - \frac{r_-}{\ell^2} (\Delta t) \right] \\
	&- \sqrt{(\alpha(r)-1)(\alpha(r')-1)} \cosh\! \left[\frac{r_+}{\ell^2}(\Delta t)  - \frac{r_-}{\ell}( \Delta \phi - 2 \pi n)  \right],   
\end{align}
where $\alpha(r) = \frac{r^2 - r_-^2}{r_+^2 - r_-^2}$, $\Delta t = t-t'$, and $\Delta \phi = \phi-\phi'$.

Note that the $n=0$ term in the image sum representation of the BTZ Wightman function (\ref{eq:rotating-btz-wightman}) corresponds to the vacuum Wightman function of AdS spacetime.

\subsubsection{Temperature of the quantum field}

While classical spacetimes might not have any associated temperature, Hawking and Unruh showed that when accounting for quantum effects, the curvature of spacetime can generate a notion of temperature \cite{Hawking_1975_Particle-creation-by-BH,Unruh_1976_Notes-on-BH-evaporation}.

In particular, the celebrated Hawking radiation has led to important developments in  gravitational physics, notably including black hole thermodynamics \cite{Bekenstein_1973_BHs-and-entropy,Bardeen-Carter-Hawking_1973_4-laws-BHs} and the black hole information paradox \cite{Hawking_1975_Particle-creation-by-BH,Hawking_1976_Breakdown-of-predictability}.

More generally, we are able to describe the thermal properties of a quantum field in curved space time via the KMS temperature \cite{Kubo_1957,Martin-Schwinger_1959}. This generalization of the Gibbs temperature describes the state of a field that is in thermodynamic equilibrium.

The KMS temperature, $T$, is related to the Hawking temperature, $T_H=\frac{1}{2\pi\ell^2}\frac{r_+^2-r_-^2}{r_+}$ \cite{Carlip_1995_BTZ}, by 
\begin{align*}
    T &= \frac{T_H}{\gamma} = \frac{1}{2\pi\ell}\frac{r_+^2-r_-^2}{r_+} \frac{r_+}{\sqrt{(r^2-r_+^2)(r_+^2-r_-^2)}} \\
    &= \frac{1}{2\pi\ell} \left(\frac{r_+^2-r_-^2}{r^2-r_+^2}\right)^{1/2} \\
    &= \frac{1}{2\pi\ell}(\alpha(r)-1)^{-1/2}.
\end{align*}
Note that the KMS temperature $T$ goes like $1/r$. So a fixed radius implies fixed KMS temperature. In particular, these can be related by $\alpha(r)-1=(4\pi^2T^2\ell^2)^{-1}$.

\subsection{Unruh-DeWitt particle detectors}
\label{sec:UDW-detector}

While the notion of measurement has been the cause of quite some consternation in quantum mechanics, the measurement problem takes a different flavour in QFT. 
Since a quantum field is a non-local object, we cannot measure it directly in a physical sense. Such a procedure would require us to perform the measurement instantaneously across all of space.
Instead, we can access information about the global state of a quantum field by performing a measurement on a detector that is locally coupled to the field.

The most common instance of this is the Unruh-DeWitt (UDW) detector \cite{Unruh_1976_Notes-on-BH-evaporation,DeWitt_1979}: a two-level quantum detector which is known to effectively model the light-matter interaction without exchange of angular momentum \cite{Lopp-MM_2021_LMI-RQI}.

The UDW detector has ground state, $\ket{0_D}$, and excited state, $\ket{1_D}$, separated by an energy gap, $\Omega$. The detector travels along a trajectory $x(\tau)$ parametrized by its proper time, $\tau$. It is coupled to a massless scalar field, $\hat{\phi}$, with the coupling described by the interaction Hamiltonian
\begin{equation}
\label{eq:udw-interaction-hamiltonian}
	H_I = \tilde{\lambda}  \chi(\tau) \left( e^{i\Omega\tau} \sigma^+ + e^{-i\Omega\tau} \sigma^- \right) \otimes \hat{\phi}[x(\tau)],
\end{equation}
where $\lambda = \tilde{\lambda} \sqrt{\ell}$ is the dimensionless coupling constant, and $\sigma^+ = \ket{1_D}\bra{0_D}$ and $\sigma^- = \ket{0_D}\bra{1_D}$ are ladder operators. 
The term containing the ladder operators can alternately be written as $\hat{\sigma}_x(\tau)$, with the time evolution generated by the detector's free Hamiltonian 
\begin{equation}
\label{eq:hamiltonian-UDW}
    \hat{H}_D=\frac{\Omega}{2}(\hat{\sigma}_z+{1}),
\end{equation}
where $\hat{\sigma}_x$, with $\hat{\sigma}_z$ are the standard Pauli spin operators acting on the detector's Hilbert space. 

In general, the switching function, $\chi(\tau)$, characterizes the strength and duration of time of the detectors coupling to the field. However, since we will be working in the framework of open quantum system, we set $\chi(\tau)=1$.

Within perturbation theory, to leading order in $\lambda$, the response function
\begin{equation}
\label{F-response}
	F(\Omega) = \int_{-\infty}^\infty d\tau \int_{-\infty}^\infty d\tau'\, e^{-i\Omega(\tau-\tau')}\, W(x(\tau),x(\tau')),
\end{equation}
is proportional to the transition probability from $\ket{0_D}$ to $\ket{1_D}$, where 
\begin{equation}
    W\left(x(\tau),x(\tau')\right) = \bra{0}\hat{\phi}(x(\tau))\hat{\phi}(x(\tau'))\ket{0}.
\end{equation}
is the Wightman function describing the field correlations between the points $x(\tau)$ and $x(\tau')$.

Given a stationary trajectory, the Wightman function depends only on the difference in times $\Delta\tau=\tau-\tau'$. For such trajectories, we can define the response rate
\begin{equation}\label{eq:response-rate}
	\F(\Omega) = \int_{-\infty}^\infty d\Delta \tau \; e^{-i\Omega\Delta\tau}\, W(\Delta\tau),
\end{equation}
which can also be understood as the derivative of the response function per unit time.

\subsection{Fisher Information}\label{sec:FI}

The Fisher information is a statistical quantity that characterizes how sensitive some observable parameter $x\in X$ is to fluctuations of an underlying parameter $\xi\in\Xi$, where the relationship between these parameters is described by the probability distribution $p(x|\xi)$, and $(X,\Xi)$ corresponds to the set of possible observable and underlying parameter values.

The Fisher information, $\I(\xi)$, is defined to be 
\begin{equation}
\label{eq:fisher-info-def}
    \I(\xi) = \int p(x|\xi) \left( \D{\ln p(x|\xi)}{\xi} \right)^2 dx = \int \frac{1}{p(x|\xi)} \left( \D{p(x|\xi)}{\xi} \right)^2 dx.
\end{equation}
This is an integral (or sum in the discrete case) of the square of the logarithmic derivative of this probability with respect to the underlying parameter over all possible observable values, weighted by the probability distribution $p(x|\xi)$.

In the information theoretic realm, the Fisher information can be related to other physically relevant values such as the Kullbeck-Leibler divergence, however its main appeal comes from its importance in the Cramer-Rao bound.

Given the unbiased estimator, $\hat{\xi}: X^n \to \Xi$, which maps $n$ observed parameter values to the underlying parameter $\xi$, the Cram\'er-Rao bound \cite{Rao_1992,Cramer_1999}
\begin{equation}\label{eq:crbound}
	\text{var}(\hat\xi) \geq \frac{1}{n\I(\xi)},
\end{equation}
imposes a lower bound on the variance of the estimator that is inversely proportional to the Fisher information. In experimental contexts, this bound and the Fisher information, allow for optimal experimental design. 

We note that while the number of observable parameter measurements, $n$, is also inversely proportional to the theoretically optimal variance in the Cramer-Rao bound, it does not factor into the Fisher information that we calculate.

\section{Derivations} \label{sec:derivations}

\subsection{Metrology with Unruh-DeWitt detectors}
\label{sec:Metrology}

 We will make use of an open quantum systems (OQS) framework \cite{BreuerPetruccione_2007_OQS} to describe the dynamics of our problem. This approach is generally effective when the system is much smaller than the environment, so it can be assumed that the system does not affect the dynamics of the environment. While there have been some recent results exploring alternative OQS frameworks for UDW detectors \cite{chen2024quantumfisherinformationcosmic,han2025reliablequantummasterequation} in response to questions surrounding the regimes of validity of various approximations \cite{kaplanek_2023}, we will consider here the standard approach, noting that the qualitative behaviour of the Fisher information appears to be largely unchanged when relaxing the assumptions \cite{chen2024quantumfisherinformationcosmic}.

Here, the detector's system is denoted by the subscript $D$, while the quantum field's system is denoted by the subscript $\phi$. The total Hamiltonian can then be expressed as
\begin{equation}
    H = H_D + H_\phi + H_I,
\end{equation}
where $H_\phi= \frac{dt}{d\tau} \sum_\mathbf{k} \omega_\mathbf{k} a_\mathbf{k}^\dagger a_\mathbf{k}$ is the free Hamiltonian of massless scalar field $\hat{\phi}$, with the redshift factor $\frac{dt}{d\tau}$ allowing the total Hamiltonian to describe time evolution in terms of the detector's proper time.
The remaining Hamiltonian terms remain as they were defined in the previous Section with $H_D$ is the free Hamiltonian of detector defined in Equation (\ref{eq:hamiltonian-UDW}), and $H_I$ is UDW detector's interaction Hamiltonian from Equation (\ref{eq:udw-interaction-hamiltonian}).

The dynamics of the total system, $\rho_\text{tot}$, are given by the von Neumann equation
\begin{equation}
    \D{\rho_\text{tot}}{\tau} = -i [H, \rho_\text{tot}].
\end{equation}
If we assume that the initial state $\rho_\text{tot}(0) = \rho_D(0) \otimes \ket{0_\phi}\bra{0_\phi}$ is separable with $\rho_D(0)$ being the initial state of the detector and $\ket{0_\phi}$ is conformal vacuum of $\hat{\phi}(x)$, we can obtain the time-evolved state of the detector by taking the partial trace over the field of the combined state, i.e., $\rho_D = \Tr_\phi \rho_\text{tot}$. 
 
If we assume a weak coupling ($\lambda \ll 1$) with field correlations decaying sufficiently fast for large time separations, then the density operator of the detector's evolution is expressed by the master equation of Kossakowski-Lindblad form 
\cite{Benatti-Floreanini_2004}. This is the most general description of Markovian time evolution of a quantum system \cite{Manzano_2020}, and is given by
\begin{equation}\label{eq:lindblad}
    \D{\rho_D(\tau)}{\tau} = -i[H_\text{eff}, \rho_D(\tau)] + L[\rho_D(\tau)],
\end{equation}
where $H_\text{eff} = \frac12\Tilde{\Omega}(\ket{0_D}\bra{0_D} - \ket{1_D}\bra{1_D})$ is the effective Hamiltonian, and
\begin{equation}
    L[\rho] = \frac12 \sum_{i,j=1}^3 C_{ij} \left( 2\sigma_j\rho\sigma_i - \sigma_i\sigma_j\rho - \rho\sigma_i\sigma_j\right),
\end{equation}
where the $\sigma_i$ are the Pauli matrices. 

The quantity $\Tilde{\Omega}$ is a renormalized gap given by 
\begin{equation}
    \Tilde{\Omega} = \Omega + i \left[ \mathcal{K}(-\Omega) - \mathcal{K}(\Omega) \right],
\end{equation}
where $\mathcal{K}(\Omega)$ is the Hilbert transform of the response per unit time $\F(\omega)$ defined by
\begin{equation}
    \mathcal{K}(\Omega) = \frac{1}{i\pi} \text{ PV} \int_{-\infty}^\infty d\omega \, \frac{\F(\omega)}{\omega - \Omega},
\end{equation}
with PV denoting the Cauchy principal value. $C_{ij}$ is called the Kossakowski matrix, and is also completely determined by the response per unit time $\F(\Omega)$:
\begin{equation}
    C_{ij} = \begin{pmatrix}
    A & -iB & 0 \\
    iB & A & 0 \\
    0 & 0 & A+C
    \end{pmatrix},
\end{equation}
where
\begin{equation}\label{eq:A}
    A = \frac12[\F(\Omega) + \F(-\Omega)],
\end{equation}
\begin{equation}\label{eq:B}
    B = \frac12[\F(\Omega) - \F(-\Omega)],
\end{equation}
\begin{equation}
    C = \F(0) - A.
\end{equation}

Eq. \eqref{eq:lindblad} can be solved analytically. 
Given a detector initialized in the general pure state $\ket{\psi_D} =
\cos\frac{\theta}{2} \ket{0_D} + \sin\frac{\theta}{2} \ket{1_D}$, its density matrix at time $\tau$ is specified by the Bloch vector $\vec{a} = (a_1,a_2,a_3)$ such that
\begin{equation}\label{eq:det-state}
    \rho(\tau) = \frac12 \left( I + \vec{a}(\tau) \cdot \vec{\sigma}  \right),
\end{equation}
where $\vec{\sigma} = (\sigma_1,\sigma_2,\sigma_3)$ are the Pauli matrices, and the Bloch vector components are given by
\begin{equation}
    a_1(\tau) = e^{-A\tau/2} \sin\theta\cos\Tilde{\Omega}\tau,
\end{equation}
\begin{equation}
    a_2(\tau) = e^{-A\tau/2}\sin\theta\sin\Tilde{\Omega}\tau,
\end{equation}
\begin{equation}\label{eq:az}
    a_3(\tau) = -e^{-A\tau} \cos\theta - R(1-e^{-A\tau}).
\end{equation}
where $R = B/A$. Note that $|\vec{a}| < 1$ in general, implying that the evolution is non-unitary.

Having identified our state of interest and its time evolution dynamics, we are now in a position to compute the Fisher information for estimating temperature for the rotating BTZ black hole with UDW detectors. Our estimation strategy is to first let the detector interact with a massless scalar field in the spacetime of interest then make a projective measurement of the detector's state after some detector proper time $\tau$.

Since we are working with a two-level quantum detector, the UDW detector has associated to it a 2-dimensional Hilbert space. It then follows that every measurement has two possible outcomes. By definition of the Fisher information, Eq. \eqref{eq:fisher-info-def}, our generally continuous probability distribution $p(x|\lambda)$ reduces to the discrete probability of getting either outcome, and the integral simplifies to a two term sum.
 
If said outcomes occur with probabilities $p$ and $1-p$ then the Fisher information can be expressed as
\begin{equation}
    \I(\lambda) = \frac{1}{p} \left( \D{p}{\xi} \right)^2 + \frac{1}{1-p} \left( -\D{p}{\xi} \right)^2 = \frac{1}{p(1-p)} \left( \D{p}{\xi} \right)^2.
\end{equation}

When measuring the detector in a state specified by the Bloch vector $\vec{a}$ using the computational basis, $\{ \ket{0_D}, \ket{1_D} \}$, the probability of obtaining $\ket{0_D}$ is given as
\begin{equation}
    p = \Tr(\rho\ket{0_D}\bra{0_D})  = \frac12(1+a_3),
\end{equation}
and so the probability of obtaining $\ket{1_D}$ is $1-p = \frac12(1-a_3)$. 

We can then express the Fisher information as
\begin{equation}\label{eq:fisher}
    \I(\xi) = \frac{(\partial_\xi a_3)^2}{1-a_3^2},
\end{equation}
where if we set the underlying parameter $\xi$ to be $T$, we find our desired (thermal) Fisher information
\begin{equation}\label{eq:fisher-t}
    \I(T) = \frac{(\partial_T a_3)^2}{1-a_3^2}
\end{equation}
In order to make this value unitless, we rescale $\I(T)$ by $T^2$ and define $\I=\I(T)T^2$ which we will hence forth simply refer to as the Fisher information.

\subsection{Response Rate for a Co-Rotating Detector}

Since the detector dynamics relevant to our calculation of the Fisher information are fully encoded in the response rate, we will work out its specific expression for our set-up.

Previously, we saw that the response rate can be understood as the response of the detector per unit time.
In the case of a co-rotating detector in BTZ spacetime,
by performing the substitution $\Delta\tau=\frac{1}{2\pi T} (z-2\pi n r_-/\ell)$, we can rewrite the response rate (\ref{eq:response-rate}) as
\begin{equation*}
    \F_{\rm BTZ} = \frac{1}{4\pi\sqrt{2}} \sum_{n=-\infty}^\infty \eta^n \int_{-\infty}^\infty d{z} ~ \left[ \frac{e^{-\frac{i\Omega}{2\pi T}(z-\frac{2\pi n r_-}{\ell})}}{\sqrt{\alpha^-_n - \cosh(z)}} - \zeta \frac{e^{-\frac{i\Omega}{2\pi T}(z-\frac{2\pi n r_-}{\ell})}}{\sqrt{\alpha^+_n - \cosh(z)}} \right]
\end{equation*}
where
\begin{align}
	\alpha^\mp_n &=  \left(1 + 4 \pi^2 \ell^2 T^2 \right)\cosh (2 \pi n \rp/\ell)  \mp 4 \pi^2 \ell^2 T^2 \, .
\end{align}

Computing the integrals, we can rewrite the response rate as
\begin{align} 
\label{eq:rBTZ-response-rate}
	\F_{\rm BTZ} &= \frac{1}{4} \left[1 - \tanh\left(\frac{\Omega}{2 T}\right) \right] \sum_{n=-\infty}^{n = \infty} e^{\frac{i\Omega n r_-}{\ell T}} \left[P_{-\frac{1}{2} + \frac{i \Omega}{2 \pi T}} \left( \alpha_n^- \right) - \zeta  P_{-\frac{1}{2} + \frac{i \Omega}{2 \pi T}} \left( \alpha_n^+ \right) \right].
\end{align}
where $P_{\nu}$ is the associated Legendre function of the first kind, satisfying $P_{-1/2 + i \lambda} = P_{-1/2 - i \lambda}$, and $T$ is the KMS temperature of the vacuum. 

This, in concert with the Fisher information results from Section \ref{sec:Metrology} allows us to compute the desired results.

\section{Results}\label{sec:results}

The thermal Fisher information of a UDW detector was first examined for 4-dimensional dS and AdS spacetimes \cite{DuMann_2021}. There, a class of 5 distinct temporal behaviours were noted in which the Fisher information starts at zero and eventually asymptotes to a non-zero finite value. 

When considering the same set up in one fewer dimensions, additional temporal behaviours become possible.
In the case of 3-dimensional AdS spacetimes, there are an additional 3 distinct classes of behaviours that can be exhibited by the Fisher information. The class of behaviours observed for 3-dimensional static BTZ spacetimes is the same as for 3-dimensional AdS spacetimes \cite{PattersonMann_2023}. We show here that this class of 8 distinct behaviours remains maximal for more generic BTZ spacetimes accounting for angular momentum.

The progression from dS to AdS to BTZ spacetimes can be seen as a series of generalizations from the perspective of the Fisher information, 
with each additional spacetime enabling, in principle a larger class of temporal behaviours. 
A comoving detector in dS experiences the same thermal response as an accelerated detector in AdS with transparent boundary condition \cite{DuMann_2021}, while the Neumann and Dirichlet boundary conditions present in AdS enable new behaviours. 
Then, the response of the accelerated detector in AdS corresponds to the $n=0$ term of the image sum in the stationary detector in a static BTZ spacetime's response.

In this paper, we bring this generalization to its apex by considering the Fisher information of a co-moving UDW detector in a rotating BTZ black hole spacetime, where there is an additional parameter corresponding to the angular momentum of the black hole.

\subsection{Temporal behaviours of the Fisher information}

\begin{figure}[htbp]
  \centering
  \includegraphics[width=\textwidth]{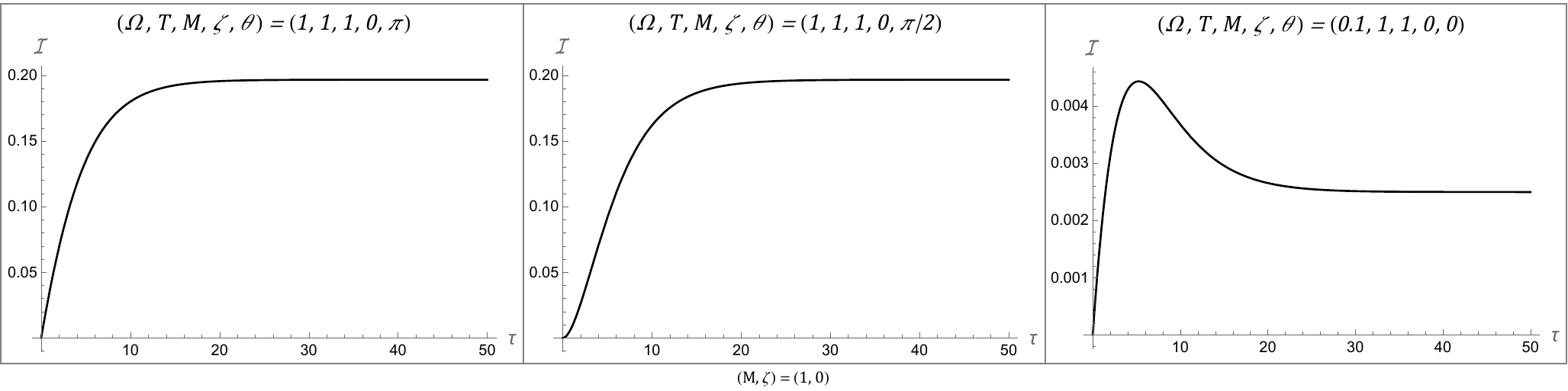}
  \caption{The first three temporal behaviours, which we label as Behaviours 1 to 3, display the most simple time-evolution of the Fisher information. The first behaviour consists of monotonically decreasing growth until it asymptotes, while the second behaviour has a point of inflection at early-time before following the same trend. Behaviour 3 achieves a global maximum at early-time before asymptoting to a finite value. All of the figures were plotted with $M=1$, and $\zeta=0$.}
  \label{fig:rBTZ-FI-Behaviours-1-through-3}
\end{figure}

\begin{figure}[htbp]
  \centering
  \includegraphics[width=0.7\textwidth]{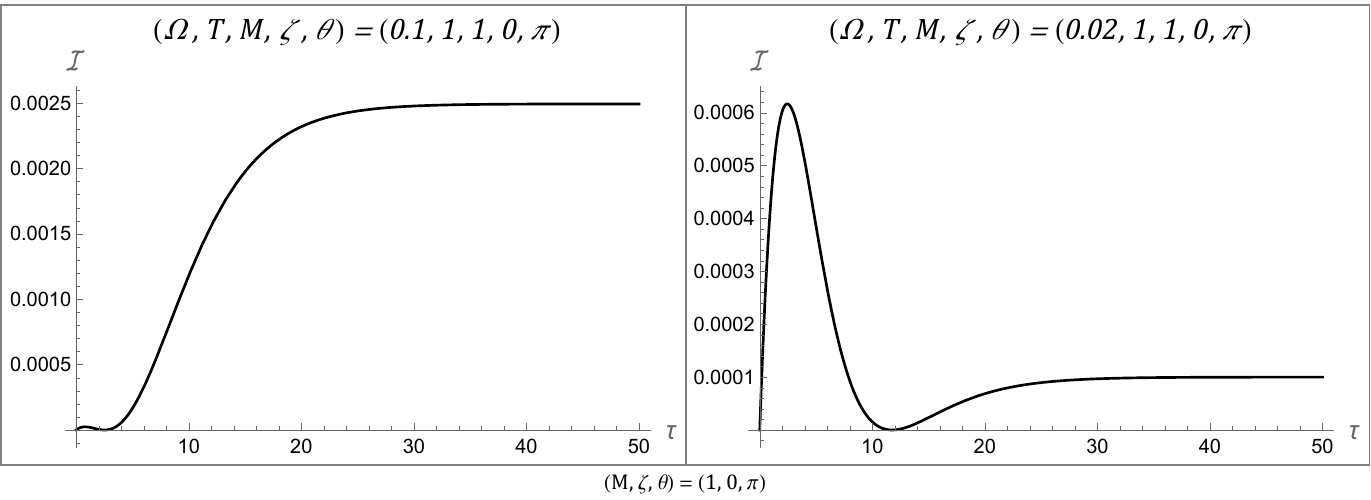}
  \caption{Temporal Behaviours 4 and 5 are similar to Behaviour 3 in that they have a local maximum at early time, however they also have a local minimum that attains $\I=0$ at an intermediary time before asymptoting to a finite value. All of the figures were plotted with $M=1$, $\zeta=0$, and $\theta=\pi$.}
  \label{fig:rBTZ-FI-Behaviours-4-and-5}
\end{figure}

\begin{figure}[htbp]
  \centering
  \includegraphics[width=\textwidth]{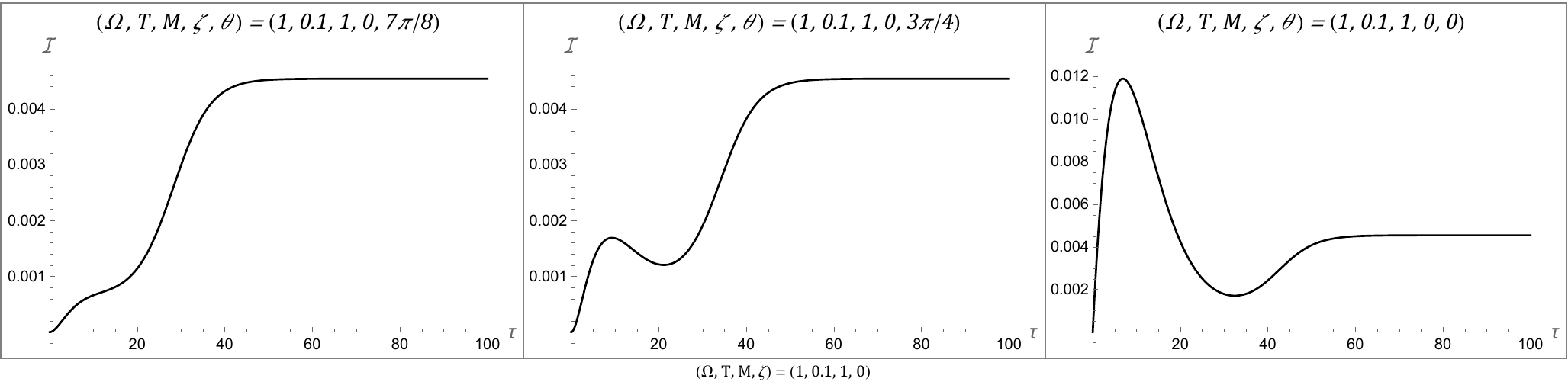}
  \caption{The final three temporal behaviours, which we label as Behaviours 6 to 8, resemble Behaviours 4 and 5 in that they have two inflection points. Behaviours 7 and 8 have a local maximum, followed by a local minimum at a non-zero value before asymptoting to a finite value, while Behaviour 6 displays monotonic growth (with two inflection points) up until it asymptotes. All of the figures were plotted with $M=1$, and $\zeta=0$.}
  \label{fig:rBTZ-FI-Behaviours-6-through-8}
\end{figure}

Examining the temporal behaviours displayed by the Fisher information for a rotating BTZ black hole, we recover the class of 8 behaviours that were present for 3-dimensional AdS and static BTZ spacetimes. 
These behaviours are plotted in Figures \ref{fig:rBTZ-FI-Behaviours-1-through-3} through \ref{fig:rBTZ-FI-Behaviours-6-through-8}, with all spacetimes chosen for simplicity to have a transparent boundary condition ($\zeta=0$) and a BTZ mass of unity ($M=1$). 
Since there were no additional behaviours enabled by non-zero angular momentum, we chose $J=0$ across all of these plots. 
That said, a variation in angular momentum can result in significant quantitative changes in the Fisher information, as we will see in the following section.
All of these behaviours remain accessible for both Neumann and Dirichlet boundary conditions, as well as for mass $M\neq1$.
The consistency of the rotating BTZ spacetime's results with previous results suggests that this class of 8 functions might be complete in the sense that it encompasses all possible temporal behaviour for the Fisher information.

It is understood why the Fisher information would have an initial value of zero: at zero time, the detector would not have had the chance to interact with the system and would thus not be able to convey any information about the system upon measurement. Similarly, the asymptotic behaviour at late time can be understood as the detector reaching a state of equilibrium with the field, thus limiting the information available at late time to some finite non-zero amount.

This asymptotic Fisher information is fully characterized by the detector's energy gap and the temperature of the field, given by
\begin{equation}
    \lim_{\tau\to\infty} \I = \frac{\Omega^2}{4T^2}{\rm sech}^2\left(\frac{\Omega}{2T}\right).
\end{equation}

The more interesting behaviour is what happens between these two stages.
For one, we can observe large peaks at an intermediary time, as seen in Behaviours 3, 5, and 8. The time at which this peak in the Fisher information occurs corresponds to the time at which one should measure the detector to optimally estimate the field temperature. 
On the other hand, there are also instances in which following such a peak there is dip in the Fisher information. These dips can actually be vanishingly small, as seen in Figure \ref{fig:rBTZ-FI-Behaviours-4-and-5}. The fact that the Fisher information can be zero at this intermediary time tells us that no information about the temperature can be extracted from the system at this time. This second (sometimes global) minimum is not well understood from a physics perspective and could warrant further investigation.

\subsection{Fisher information under varying angular momentum}

Having characterized the general temporal Fisher information behaviour, we turn our attention to its sensitivity to variations of the black hole parameters. That is, we consider how the spacetime parameters influence the Fisher information behaviour, and how tuning our detector might accentuate or attenuate these effects.
%further affect these behaviours.

Previous work has shown that the Fisher information can be sensitive to the mass of the black hole for the appropriate boundary conditions \cite{PattersonMann_2023}.
And while mass is the parameter associated with the simplest of black holes, more realistic black hole models must account for rotation.
While this is especially true in the context of astrophysical black holes, where almost all known black holes have non-negligible angular momentum, rotating black holes are known to result in distinctive behaviours in the context of RQI \cite{Robbins_2020_Ent-amp-rBTZ, Robbins_2021_Anti-Hawking-rBTZ}. 
As such, the following analysis of the Fisher information for a rotating BTZ black hole extends our understanding of the Fisher information's applicability to relativistic quantum metrology, while also helping paint a fuller picture of the rotating BTZ spacetime's role in RQI.

Algebraically, the difference between the response rate of a co-rotating detector in a rotating BTZ spacetime and the static detector in a static BTZ spacetime can be constrained to the exponential multiplicative factor of $\exp(i\Omega n\rn/\ell T)$ in the image sum (\ref{eq:rBTZ-response-rate}).
Indeed, when the inner radius is vanishing, i.e., when $\rn=0$, the response rate of the co-rotating detector given by Eq. (\ref{eq:rBTZ-response-rate}) reduces to that of a static detector outside a static BTZ black hole \cite{PattersonMann_2023}.
Since $J\sim\rn$, we have that for a fixed black hole mass, adding a small amount of angular momentum is not likely to perturb the system significantly. However, we expect that for systems with near extremal angular momentum, there may be significant perturbations of the Fisher information.

In order to better describe the Fisher information in the context of an extremal black hole, i.e., when $J=M\ell$, we will express the angular momentum $J$ of a BTZ black hole as a fraction of its BTZ mass $M$. 

We note that in order to produce the plots of the Fisher information, the image sum for Eq. (\ref{eq:rBTZ-response-rate}) was truncated to a finite number of terms. The number of terms included varied depending on the other parameters from our set-up, with smaller masses requiring a larger number of terms, while large masses required fewer image sum terms to achieve numerical convergence.

\begin{figure}[ht!]
	\centering
	\includegraphics[width=\linewidth]{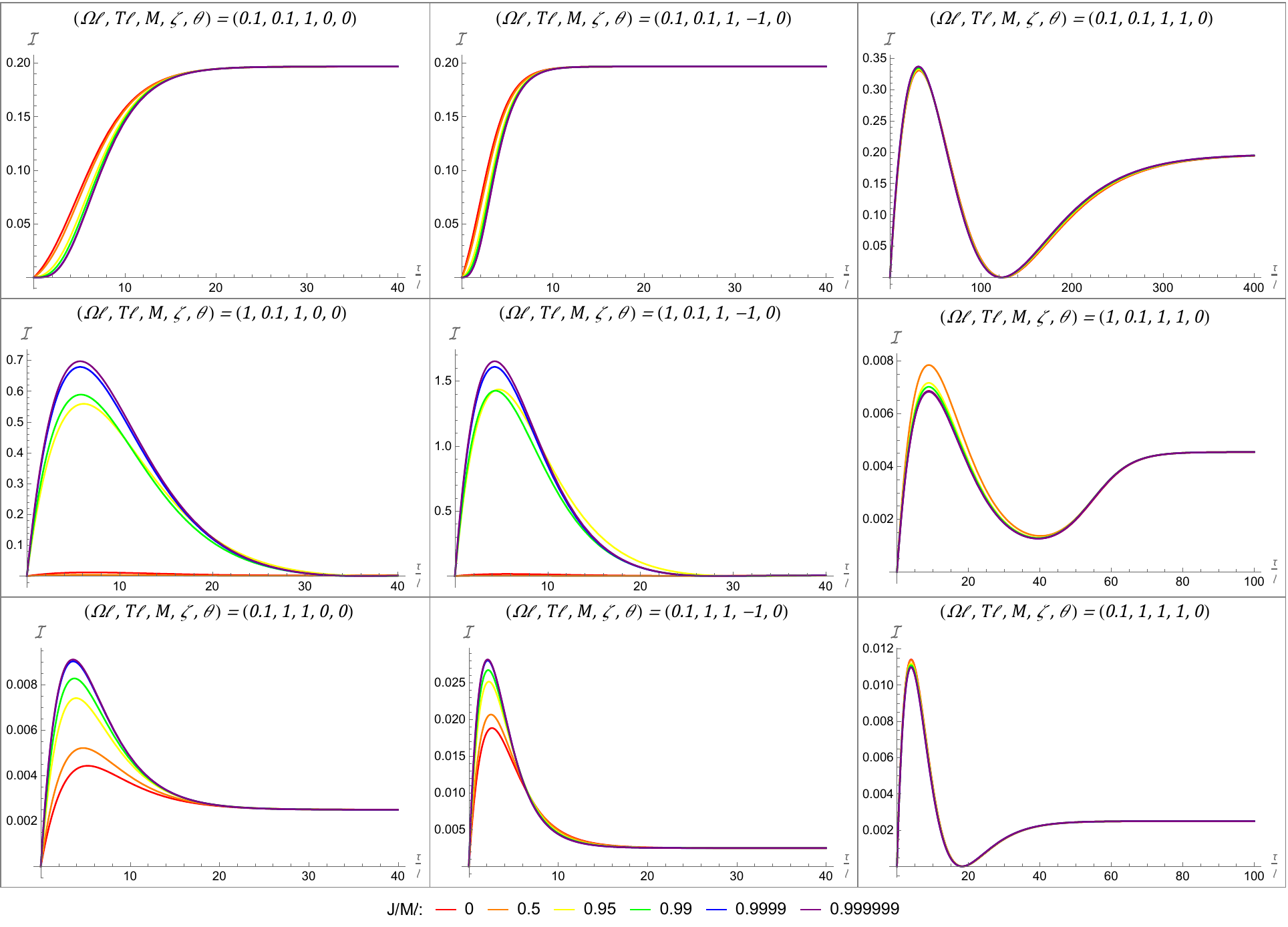}
	\caption[Fisher information in the rotating BTZ spacetime for $M=1$ and varied angular momentum]{These nine plots demonstrate the effect that angular momentum has on the Fisher information for a black hole with a large mass $M=1$.
Each column represents a different boundary condition with the left-to-right order being: transparent $(\zeta=0)$, Neumann $(\zeta=-1)$, and Dirichlet $(\zeta=1)$ boundary conditions. 
Each row represents a different energy gap to temperature ratio, with the top-to-bottom order being: neutral $(\Omega/T = 1)$, `cold' $(\Omega/T = 10)$, and `hot' $(\Omega/T = 1/10)$.
Within a given plot, each curve represents a different angular momentum, corresponding to increasing angular momentum (from $J/M=0$ to $J/M=0.999999$) as we move along the colours of the rainbow, from red to purple.}
	\label{fig:rbtz-theta0m1}
\end{figure}

The effect of the black hole's rotation on the Fisher information is first depicted in Figure \ref{fig:rbtz-theta0m1} for a black hole with a large mass of $M=1$. In each plot, we have considered a range of angular momenta varying from the static case ($J/M\ell=0$ in red) to an almost maximally rotating black hole ($J/M\ell=0.999999$ in purple) using the colours of the rainbow.
The grid layout of Figure \ref{fig:rbtz-theta0m1} allows us to show how the the Fisher information varies when considering the energy gap-to-temperature ratio (held constant in a given row) and the boundary condition (held constant in a given column).
Throughout Figure \ref{fig:rbtz-theta0m1}, we have initialized the detector in the state parametrized by $\theta=0$ and have fixed the BTZ mass to be $M=1$.

We begin by noting that given a fixed angular momentum (i.e., looking at a single colour of lines) the behaviour of the Fisher information for the transparent and Neumann boundary conditions are qualitatively nearly identical, while the Dirichlet boundary condition is qualitatively quite different.
This similarity between the transparent and Neumann boundary conditions, and their difference from the Dirichlet boundary condition is in agreement with other RQI results for both static and rotating BTZ spacetimes \cite{Robbins_2020_Ent-amp-rBTZ,Robbins_2021_Anti-Hawking-rBTZ,Henderson_2020_Anti-Hawking-BTZ, Hodgkinson_2012_Static, suryaatmadja_2023_signatures-rotating-BTZ,PRNML_2024_MoreExcitement,wang_2024_singular-excitement-rBTZ}.
That said, accounting for the Fisher information seems to further accentuate this discrepancy. While there is a clear qualitative difference between curves of the same color (corresponding to a fixed value of $J/M\ell$) in Figure \ref{fig:rbtz-theta0m1}, the variation of the Fisher information between angular momenta can also be used as a differentiator. Moreover, this variation can be observed locally, and does not require global information about the temporal behaviour of the Fisher information.

When the ratio between the energy gap of the detector and the KMS temperature of the field is unity, the Fisher information is mostly unperturbed by any variations in the angular momentum of the black hole. However, for both the `cold' and `hot' configurations, there is a very dramatic change in the Fisher information induced by the black hole's rotation. 
This is most pronounced for the transparent and Neumann boundary conditions for which the maximal Fisher information can increase by more than an order of magnitude. This means that the detector is very sensitive to the angular momentum of the black hole. More interestingly, this sensitivity is most pronounced for near-extremal black holes. In the context of temperature estimation, this means that it is much easier for the detector to estimate the temperature of the field outside a rapidly rotating black hole than outside a static (or even weakly rotating) one.

However, in the case of the Dirichlet boundary condition, the same angular momentum results in a decrease in the maximal Fisher information. While not as significant of a shift, this too is most pronounced when passing from weakly rotating to near-extremal black holes.

Having shown that the angular momentum of the black hole can have a significant effect on the Fisher information for a large mass black hole, we will now turn our attention to the case a of a small mass black hole with $M=1/100$.

\begin{figure}[ht!]
	\centering
	\includegraphics[width=\linewidth]{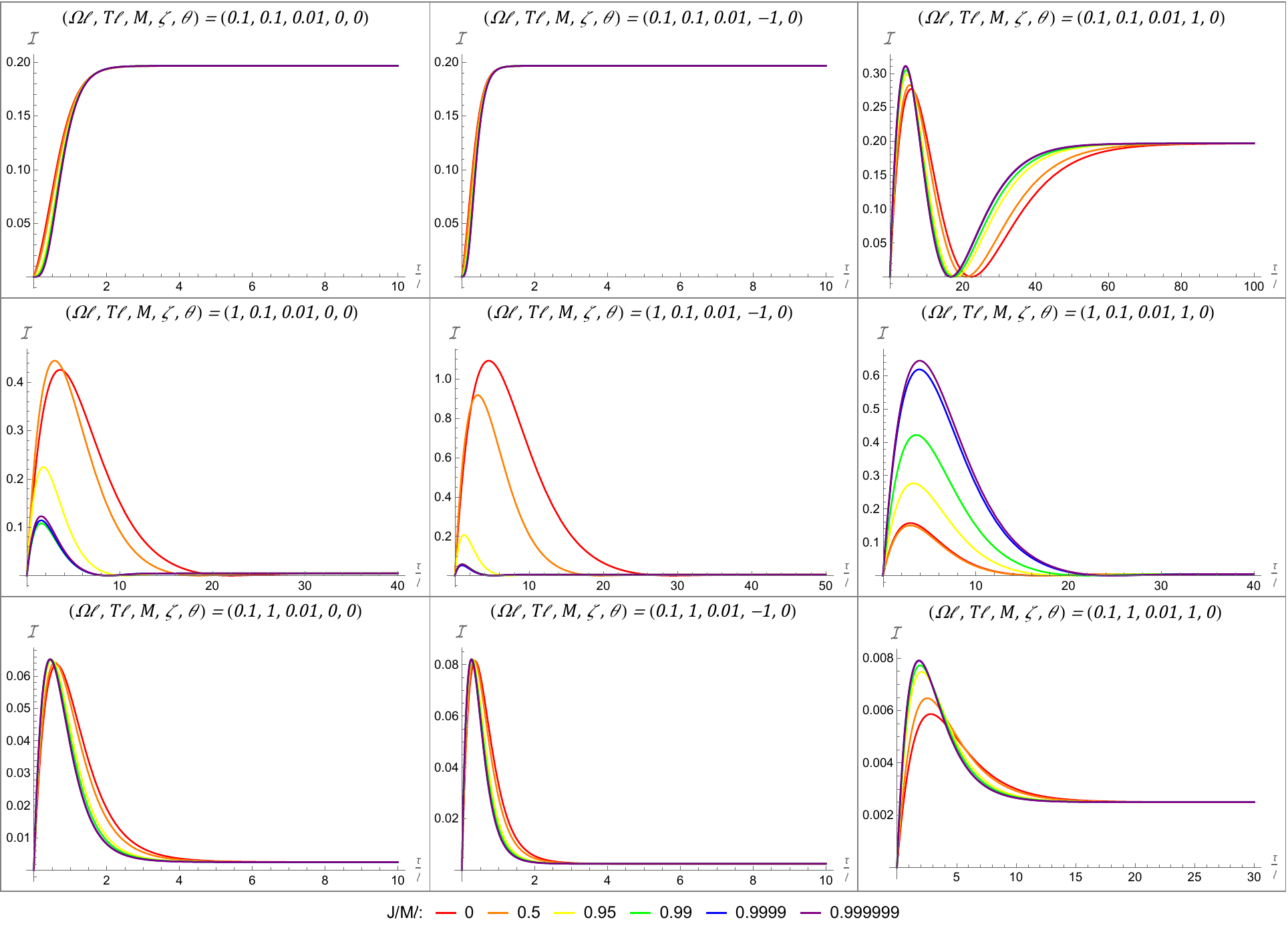}
	\caption[Fisher information in the rotating BTZ spacetime for $M=0.01$ and varied angular momentum]{Here we depict the effect of varying the angular momentum, $J$, for a small BTZ mass of $M=0.01$. While we retain some trends from the $M=1$ case depicted in Figure \ref{fig:rbtz-theta0m1}, such as not much change when $\Omega=T$, and the Dirichlet ($\zeta=1$) boundary condition having distinct behaviours compared to the other boundary condition, there are also some new effects. Most notably, for the `cold' set up, the Fisher information is now suppressed for increasing angular momentum in the transparent ($\zeta=0$) and Neumann ($\zeta=-1$) boundary condition, while there is a significant increase for the Dirichlet boundary condition.}
	\label{fig:rbtz-theta0m001}
\end{figure}

In Figure \ref{fig:rbtz-theta0m001}, we once again use a grid to convey how the Fisher information varies for different energy scales and boundary conditions. The meaning of the coloured curves remains unchanged, with the red curve corresponding to a black hole with no angular momentum ($J=0 M\ell$), while the purple curve depicts the Fisher information for a near-extremal black hole ($J=0.999999 M\ell$).

While the qualitative behaviour of the Fisher information associated to a given tile within the grid remains largely unchanged between Figure \ref{fig:rbtz-theta0m1} and Figure \ref{fig:rbtz-theta0m001}, a variation in the angular momentum of a small mass black hole can have the opposite effect than in the large mass scenario.
This effect is most pronounce when $\Omega/T=10$ (corresponding to the second row), where we the maximal Fisher information being amplified by rotation in the case of a Dirichlet boundary condition for $M=0.01$, in contrast with the decrease observed for $M=1$.
This difference in the sign of the gradient of the Fisher information as a function of the angular momentum $J$ reinforces the importance of the BTZ mass parameter $M$ previously observed \cite{PattersonMann_2023} and overlays this with the sensitivity of the Fisher information to changes in the angular momentum of the black hole.

\section{Conclusion}\label{ch:conclusion}

In this paper, we have extended the analysis of thermal Fisher information for (2+1)-dimensional black holes to account for systems with non-trivial angular momentum. 
This is an important advancement given the importance of rotation for both astrophysical black holes as well as RQI protocols.
We contextualize this work by characterizing these results as the most general analysis of thermal Fisher information for (2+1)-dimensional spacetimes. In particular, we found that all 8 of the temporal behaviours observed for the thermal Fisher information are possible for every boundary condition of the rotating BTZ spacetime.

We further showed that that the Fisher information is sensitive to the angular momentum of the black hole, with significant variance in the Fisher information for small changes in the angular momentum for near-extremal black holes. Moreover, this variation is also sensitive to the mass of the black hole.

This analysis of the thermal Fisher information serves as a contribution to both the fields of relativistic quantum metrology and to RQI. In the context of metrology, the Fisher information behaviours that we characterize in this work can be used to inform experiments (or perhaps analogue experiments) that seek to detect Hawking radiation. For RQI enthusiasts, this work improves our understanding of how UDW detectors can extract information from the quantum vacuum and what this can tell us about the structure of spacetime.

Looking ahead, there are many questions that remain to be answered. From the relativistic quantum metrology perspective, it would be interesting to extend this work explicitly to the analysis of quantum Fisher information and to also consider alternative open-quantum system frameworks \cite{chen2024quantumfisherinformationcosmic,han2025reliablequantummasterequation}. While the existing literature suggests that these extensions would lead to minor quantitative deviations from the results found here, it would be good to confirm that the deviations would not affect the qualitative results, reinforcing their robustness.

Having completed the analysis of the thermal Fisher information for (2+1)-dimensional black hole spacetimes \cite{PattersonMann_2023}, the Fisher information analysis of a Schwarzschild black hole remains to be completed. Due to the numerical nature of this type of work, increasing the catalogue of studied spacetimes is an important step in completing our understanding of these results. The asymptotically flat black hole spacetime is a notable omission from the study cosmic expansion \cite{HFZF_2018_Q-estimation-exp-spacetime,chen2024quantumfisherinformationcosmic} and accelerated detectors \cite{FengZhang_2021_QFI-Unruh}. Extensions to the Kerr class of black holes is an obvious next step in understanding the effects of angular momentum  on Thermal Fisher Information. 

Looking in a different direction, it might be interesting to consider the Fisher information of various spacetime parameters for analogue RQI set-ups and protocols \cite{Biermann_2020_analogue-Unruh,Bunney_2023_3rd-sound-unruh}. While detecting the thermal response of a detector from the Hawking radiation of an astrophysical black hole remains all but impossible, progress made in exploring RQI behaviours in analogue systems \cite{Weinfurtner:2010nu,Steinhauer:2015saa} means this might be a natural place to apply the theories of parameter estimation for black hole spacetimes.

\section*{Acknowledgements}

E.A.P. acknowledges support from an NSERC Alexander Graham Bell CGS-M scholarship.

\bibliographystyle{unsrt} 
\bibliography{references.bib}

\end{document}